\documentclass[9pt,twocolumn,twoside]{osajnl}
\journal{ol} % Choose journal (ao, aop, josaa, josab, ol, pr)

\setboolean{shortarticle}{true}
\usepackage{braket}
\usepackage{threeparttable}
\title{Broadband Fiber-based Entangled Photon-Pair Source at Telecom O-band}

\author[1,*]{Changjia Chen}
\author[1]{Calvin Xu}
\author[1]{Arash Riazi}
\author[1]{Eric Y.  Zhu}
\author[2]{Alexey V.Gladyshev}
\author[3]{Peter G. Kazansky}
\author[1]{Li Qian}

\affil[1]{Dept. of Electrical and Computer Engineering, University of Toronto, 10 King's College Rd., Toronto, M5S 3G4, Canada}
\affil[2]{Prokhorov General Physics Institute of the Russian Academy of Sciences, Dianov Fiber Optics Research Center, 38 Vavilov Street, 119333 Moscow, Russia}
\affil[3]{Optoelectronics Research Centre, University of Southampton, Southampton SO17 1BJ, United Kingdom}

\affil[*]{Corresponding author: changjia.chen@mail.utoronto.ca}
\begin{abstract}

In this letter, we report a polarization-entangled photon-pair source based on type-II spontaneous parametric down conversion at telecom O-band in periodically poled silica fiber (PPSF). The photon-pair source exhibits more than 130 nm ($\sim$24 THz) emission bandwidth centered at 1306.6 nm. The broad emission spectrum results in a short biphoton correlation time and we experimentally demonstrate a Hong-Ou-Mandel interference dip with a full width of 26.6 fs at half maximum. Owing to the low birefringence of the PPSF, the biphotons generated from type-II SPDC are polarization-entangled over the entire emission bandwidth, with a measured fidelity to a maximally entangled state greater than 95.4$\%$. The biphoton source provides the broadest bandwidth entangled biphotons at O-band to our knowledge. 
\end{abstract}

\setboolean{displaycopyright}{true}

\begin{document}

\maketitle

In the last few decades, the development of photon-pair sources has laid the foundation for quantum photonic technologies. One of the most convenient methods of producing entangled photon pairs is spontaneous parametric downconversion (SPDC) in a second-order nonlinear medium. SPDC has been extensively studied for its applications in quantum information processing and photonic quantum computation\cite{slussarenko2019photonic}. By engineering the nonlinear medium, the photon pairs generated by SPDC can be naturally endowed with nonclassical correlations, such as entanglement in polarization\cite{kwiat1995new} or in orbital angular momentum\cite{zhang2016engineering} degrees of freedom. The frequency-time correlated biphotons in SPDC sources are also of broad interest, as they enable quantum interferometry that  surpasses the limits of classical sensitivity\cite{kaiser2018quantum}. 

In particular, broad bandwidth and time-correlated photon pairs  are desirable as they enable the investigation of the optical properties of a sample over a wide spectral range, enhancing the spatial and temporal resolution of quantum sensing\cite{nasr2003demonstration}. Broadband photon-pair sources are also significant for quantum communication techniques, such as wavelength-multiplexed entanglement distribution\cite{lim2008broadband} and clock synchronization\cite{valencia2004distant}. For example, it is crucial to use broadband time-correlated photon pairs  for distant clock synchronization, as its precision is limited by the biphoton correlation time\cite{valencia2004distant}. However, in fiber-based quantum communication where the biphotons are transmitted through long-distance fiber links, precise dispersion compensation is needed because the fiber's chromatic dispersion will obscure the timing correlation. A biphoton source centered near the zero-dispersion wavelength (around 1310 nm) of the telecom silica fiber may resolve this problem and improve the precision of timing\cite{grieve2019characterizing}.
Various entangled photon sources at telecom O-band (1260 - 1360 nm) with a biphoton emission bandwidth up to 70 nm have been developed\cite{Zhong:09OE, Zhong2010, Martin:09OE, Martin2010, Liu2014, Shi2020,Xiang2020,Hall2009}. However, this bandwidth is still lacking. Generating broadband and highly time-correlated photon pairs via the process of SPDC remains a challenge. 
In order to use SPDC to generate broadband, time-correlated biphotons, various methods have been employed, such as using a short nonlinear medium\cite{lim2008broadband} and chirped quasi-phase-matching\cite{nasr2008ultrabroadband}, though at the expense of reduced biphoton generation rates and complex fabrication requirements. A time-correlated photon-pair source at O-band which has broader bandwidth and is compatible with telecom fibers is still highly desired.

In addition to the time-frequency degree of freedom, the polarization correlation within the biphotons generated via SPDC also attracts much interest. Broadband biphotons with polarization entanglement can be used for polarization sensitive quantum sensing\cite{booth2011polarization} and multi-user quantum communication\cite{Joshi2020}. However generating polarization entangled photons over a broad bandwidth often requires stable compensation for the wavelength-dependent birefringence of the nonlinear medium\cite{lim2008broadband,Zhong2010}. Interferometric methods or compensating components are therefore needed to achieve a high polarization entanglement quality\cite{kim2006phase}, increasing the experimental difficulties.

A promising solution is to use a nonlinear medium of sufficient interaction length but low birefringence and low chromatic dispersion to generate broadband entangled photon pairs . In this letter, we demonstrate a cw-pumped periodically poled silica fiber (PPSF) based entangled photon-pair source quasi-phase-matched (QPM) at 1306.6 nm, which is near the zero-dispersion wavelength of telecom single-mode silica fiber (e.g. SMF-28). The low chromatic dispersion and the low birefringence of the PPSF at telecom O-band result in a biphoton emission bandwidth greater than 130 nm ($>$ 24 THz in frequency), which to our knowledge is the broadest bandwidth of polarization-entangled biphotons at O-band. We demonstrate Hong-Ou-Mandel interference with the biphoton source and measure an interference dip width of 26.6 $\pm$ 1.3 fs. We also perform quantum state tomography to measure the quality of polarization entanglement. The concurrence\cite{wootters1998entanglement} of the polarization entanglement is measured to be more than 0.91 and the fidelity to a maximally entangled state is measured to be more than 95.4$\%$. This broadband PPSF source employs mature telecommunication components and eliminates beam alignment. It is compatible with the existing low-loss telecom fiber links and facilitates long-distance entanglement distribution and compensation-free quantum clock synchronization.

To see how birefringence and chromatic dispersion affect the SPDC spectrum, we may write the spectral brightness of an SPDC source as\cite{christ2013single}:
\begin{align}
I(\omega_s,\omega_i)\propto\mathrm{sinc}^2\left[\frac{L}{2}\left(k_A(\omega_p)-k_B(\omega_s)-k_C(\omega_i)-\frac{2\pi}{\Lambda}\right)\right],\label{eqn:spectrum}
\end{align}
where $k$ is the wavenumber of the PPSF, $L$ is the length of the PPSF, subscripts $p, s, i$ denote pump, signal and idler respectively and $\Lambda$ is the QPM period. The angular frequencies of photons obey energy conservation: $\omega_s + \omega_i = \omega_p$ and we assume $\omega_i > \omega_s$. Subscripts $A$, $B$ and $C$ represent the polarization modes	 of the photons.   In the context of this letter, we discuss the type-II SPDC in PPSF, in which a pair of orthogonally polarized photons are generated by the down-conversion of a vertically polarized pump photon, i.e. $A = V$\cite{zhu2012direct}. Therefore, in type-II SPDC, subscripts $B, C$ can be $H, V$ or $V, H$, where $H$ is for horizontally polarized photons and $V$ is for vertically polarized photons. We assume the PPSF is cw-pumped and $k_A(\omega_p) = k_V(\omega_p)\equiv k_p$, thus the QPM condition becomes:
\begin{align}
k_p - k_H(\omega_p/2) - k_V(\omega_p/2) - \frac{2\pi}{\Lambda}=0\label{eqn:QPM}.
\end{align}
Taylor-expanding the wavenumbers $k(\omega)$ in Eq. (\ref{eqn:spectrum}) about $\omega_p/2$ up to the second-order, and substituting Eq. (\ref{eqn:QPM}) into Eq. (\ref{eqn:spectrum}) to eliminate the zeroth order terms, we obtain\cite{chen2017compensation}: 
\begin{align}
I(\Delta) \propto \mathrm{sinc}^2\left[\frac{1}{2} ML\Delta + \frac{1}{2}k_2L\Delta^2+ O(\Delta^3)\right],
\end{align}
where $\Delta =\omega_p/2 - \omega_s = \omega_i - \omega_p/2$, $k_2 = \partial^2 k(\omega)/\partial \omega^2|_{\omega = \omega_p/2}$ is the chromatic dispersion at frequency $\omega_p/2$, $M = (\partial k_H(\omega)/\partial \omega - \partial k_V(\omega)/\partial \omega)|_{\omega = \omega_p/2}$ is the group velocity mismatch between H and V polarizations, and $O(\Delta^3)$ represents the terms that have orders 3 or more.  The width of $I(\Delta)$ has an inverse relation to $M$ and $k_2$. If $M$ and $k_2$ are small enough, the spectral bandwidth of SPDC can be very broad. In general, the SPDC bandwidth in second-order nonlinear crystals such as lithium niobate and potassium titanyl phosphate is significantly limited by their high group birefringence $M$ and nontrivial chromatic dispersion $k_2$. On the other hand, the group birefringence $M$ in PPSF has been shown to be negligible\cite{chen2017compensation} and the polarization entanglement of the biphotons in PPSF can be naturally generated without the need of any birefringence compensation\cite{zhu2013poled}. As a result, the bandwidth of the biphotons in PPSF sources is mainly determined by its chromatic dispersion $k_2$, and $\Delta\propto1/\sqrt{k_2L}$. Taking advantage of its low dispersion at O-band, the PPSF is able to generate entangled photons with very broad bandwidth.

\begin{figure}[t]
\centering
\includegraphics[width=8.5cm]{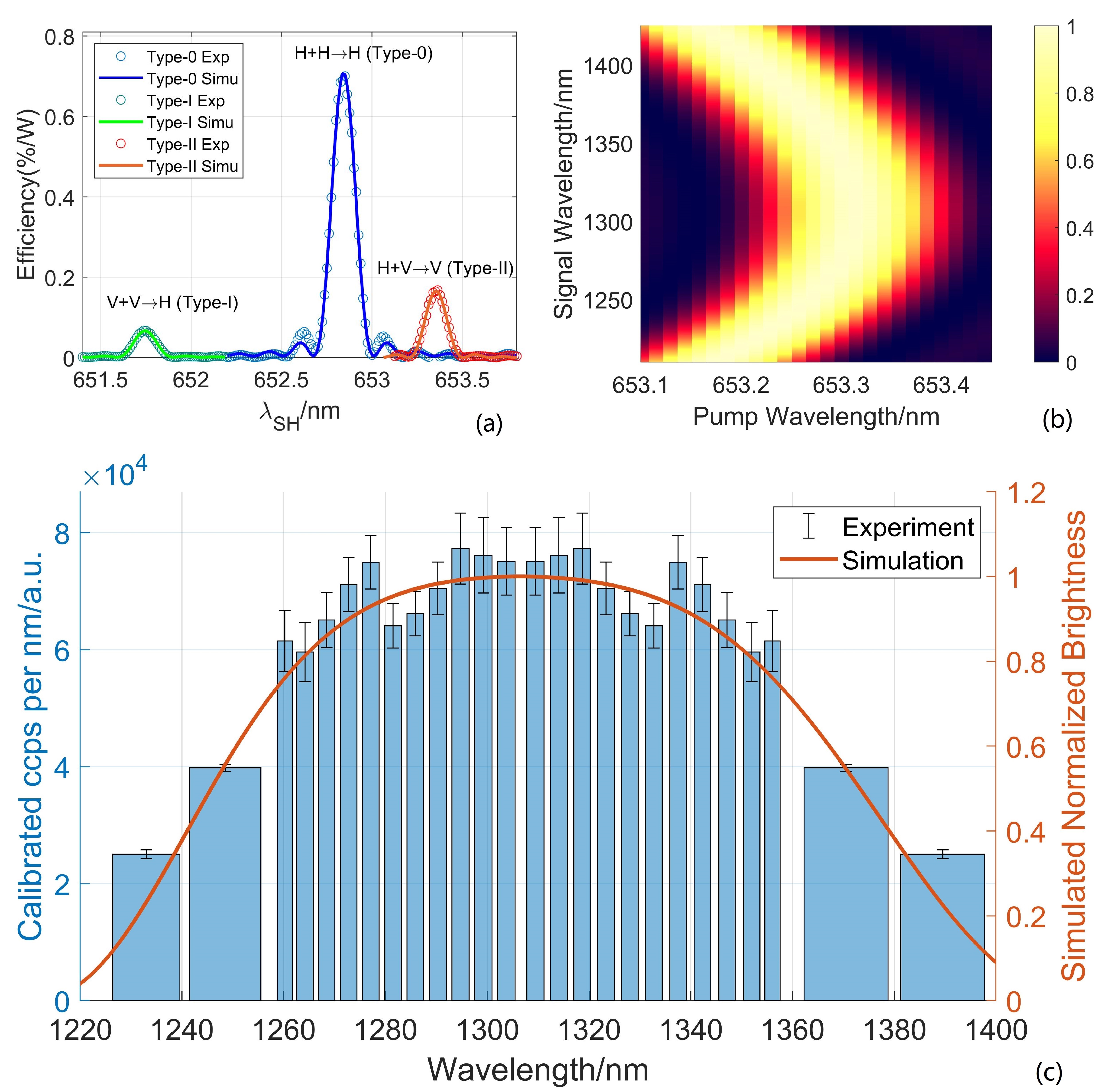}
\caption{(a) Experimentally measured and theoretically simulated SHG spectra of the PPSF. The solid-line curves have their width based on the dispersion of the fiber, while their heights are the best fit to the experimental data. The weak birefringence in PPSF results in the spectral separation of three polarization-dependent phase-matchings. (b) Theoretical tuning curve of the type-II SPDC emission spectrum in PPSF with respect to the pump wavelength. (c) Experimentally measured biphoton spectrum shows a biphoton emission bandwidth greater than 130 nm ($>$ 24 THz).}
\label{fig:SHG}
\end{figure}

The PPSF we use in the source is a 20-cm-long, weakly birefringent step-index silica fiber with an NA $\approx$ 0.14 (at the wavelength of 1.55 $\mu$m), which is compatible with telecom fibers such as SMF-28\cite{CorningSMF}. It exhibits second-order nonlinearity which is induced through thermal poling. The fabrication details of the PPSF are discussed in Ref.\cite{Canagasabey2009}. The QPM condition is achieved through periodic UV erasure with $\Lambda =54\ \mu m$. As shown by the experimentally measured second harmonic generation (SHG) spectrum in Fig.\ref{fig:SHG}(a), the PPSF supports three types of SHG, and correspondingly three types of SPDC\cite{zhu2012direct}. The conversion efficiency  is estimated to be $0.70 \%/W$ for type-0 SHG and $0.16 \%/W$ for type-II SHG. The theoretical ratio of the three peaks (Type 0, I, II) should be 9:1:4\cite{Zhu2010}. The experimental type-II peak is a bit lower than 4/9 of the type 0 peak, which can be attributed to the imprecise polarization alignment of the fundamental light in the SHG measurement. Among the three types, the type-II SPDC in PPSF can be used for compensation-free broadband polarization entanglement generation\cite{chen2017compensation}. A finite element eigenmode solver (\textit{Lumerical Inc.}) is used for calculating the birefringence and dispersion of the fiber modes based on the PPSF's geometry. Assuming a cw-pump is used,  we simulate the type-II SPDC spectrum and obtain the tuning curve as a function of the pump wavelength, as shown in Fig.\ref{fig:SHG}(b). When the pump wavelength for SPDC is set at around the type-II SHG peak (653.3 nm), the PPSF can generate broadband biphotons with more than 130 nm bandwidth.

\begin{figure*}[t]
\centering
\includegraphics[width=17cm]{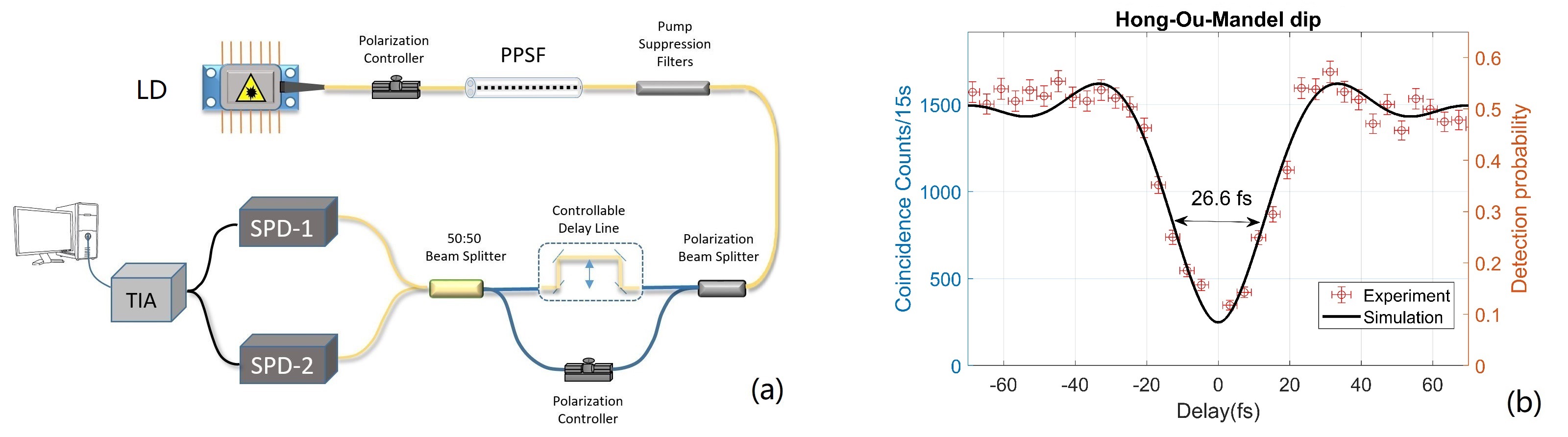}
\caption{(a) Experimental setup for the Hong-Ou-Mandel interference Measurement. LD: cw laser diode with emission wavelength at 653 nm. SPD, single photon detector (IDQ220, ID Quantique, quantum efficiency $\sim20\%$ in O-band); TIA, time interval analyzer (Hydraharp 400).  (b) Hong-Ou-Mandel interference shows a dip of 83.2 $\pm$ 0.3 $\%$ visibility with width of 26.6 $\pm$ 1.3 fs.}
\label{fig:HOM}
\end{figure*}

The biphoton spectrum of type-II SPDC in PPSF is experimentally measured with a pump laser at 653.3 nm at room temperature ($\sim22^oC$), as shown in Fig.\ref{fig:SHG}(c). To measure the biphoton spectrum, we first use a pair of fine tunable filters at O-band (\textit{O/E land Inc.}, tuning range 1260 nm -1360 nm). Outside the tuning range of the fine tunable filters, we use two coarse wavelength division multiplexers (CWDMs, each has a 3dB transmission bandwidth of 17nm) centered at 1370 nm and 1390 nm respectively. After calibrating the wavelength-dependent loss and the detector efficiencies, a spectral bandwidth of more than 130 nm (>24 THz in frequency) of type-II SPDC is obtained from the measured coincidence counts. 

In addition to the direct spectral brightness measurement, a Hong-Ou-Mandel interference (HOMI) experiment is also performed, as one may use the temporal correlation\cite{HOM1981} of the biphotons to  infer their bandwidth. The HOMI experimental setup is shown in Fig.\ref{fig:HOM}(a). A laser diode that has its wavelength stabilized at 653.3 nm by a fiber Bragg grating is used as the pump, with its polarization adjusted to align to the slow axis (defined as the V polarization in PPSF\cite{zhu2012direct}) of the PPSF at the fiber input. At the output end of the PPSF, fiber-pigtailed pump suppression filters reduce the pump power by >90 dB, at an insertion loss of 6 dB for the down-converted light. The biphotons then pass through an interferometer which is composed of a polarizing beam splitter, an inline controllable delay line (\textit{General Photonics Inc. MDL-001}), a polarization controller, and a 50:50 beam splitter. The coincidence rates are depicted in Fig.\ref{fig:HOM}(b). An HOMI dip of 26.6 $\pm$ 1.3 fs in width and $83.2\pm0.3\%$ in visibility is obtained from the fitting curve\cite{branczyk2017hongoumandel} by minimizing the error with respect to the interference pattern. Note that, the interference visibility is limited by the non-optimal components in the interferometer which are designed to operate at 1550 nm instead of the telecom O-band. Yet the dip visibility still well exceeds the classical limit. The dip width indicates that the bandwidth of biphotons is 138 nm if the biphotons are transform-limited. The dip width agrees well with our spectral measurement, which is expected because of the low dispersion in PPSF.

\begin{figure}[t]
\centering
\includegraphics[width=8cm]{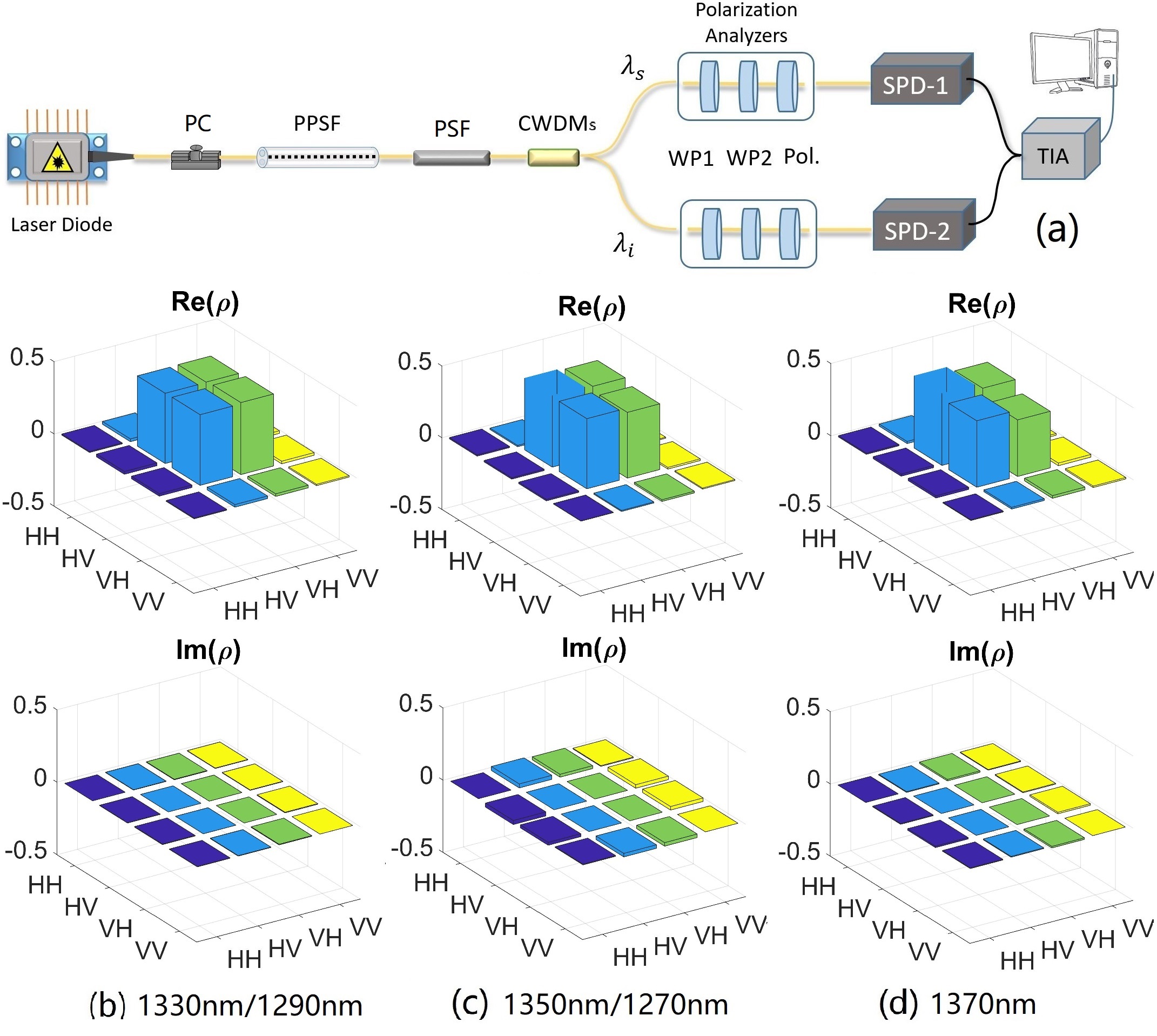}
\caption{(a) Experimental setup for polarization state tomography. PC, polarization controller; PSF, pump suppression filters; CWDMs, coarse wavelength division multiplexers; WP, waveplate; Pol., polarizer. (b)-(d) The real and imaginary part of the density matrices of the biphoton states that are measured using CWDMs centered at (b) 1330 nm and 1290 nm (c) 1350 nm and 1270 nm; (d) 1370 nm, CWDM centered on 1250nm is not available to us. Instead of using a second CWDM at 1250nm, we used the refection port of the 1370nm CWDM, which has a broad reflection spectrum, for the idler's measurement.}
\label{fig:QST}
\end{figure}

Another important and unique feature of the PPSF biphoton source is that it is capable of generating polarization entanglement over much of its bandwidth without any birefringence compensation. The polarization entanglement at various wavelengths of the emission band is measured with the quantum state tomography (QST)\cite{qstDFVJames2001PRA} experimental setup shown in Fig.\ref{fig:QST}(a). To perform QST at different conjugated wavelength pairs $\lambda_s$ and $\lambda_i$, three different CWDM sets are used as wavelength splitters. The CWDMs of 17 nm bandwidth in each set were cascaded such that conjugated wavelengths near the center wavelengths (shown in Table.\ref{Table:C}) would be transmitted. The biphoton polarization states are analyzed by two polarization analyzers (PAs, \textit{Hewlett Packard 8169A}). Each PA includes one polarizer and two waveplates. The commercial PAs were designed for 1550 nm and therefore the retardances of the waveplates were carefully calibrated for the transmission wavelengths of the CWDMs at telecom O-band. The results of the QST are summarized in Table.\ref{Table:C}, which shows the concurrences (without subtracting accidentals) of the biphoton states of each wavelength pair. The density matrices of the biphotons states are shown in Fig.\ref{fig:QST}(b)-(d). Neither the measurement apparatus nor the optical components of the source are optimal (they are all designed for 1550nm except for the PPSF), yet we are able to obtain concurrences of more than 0.91 in the biphoton emission band, even when the biphotons are more than 120 nm apart.

\begin{table}[t]
\centering
\renewcommand\arraystretch{1.4}
\caption{The experimental results of the quantum state tomography: the concurrence\cite{wootters1998entanglement} and the fidelity to a maximally polarization-entangled biphoton state that are measured using three CWDM sets at different wavelengths. Each of the CWDM filters has a top-hat transmission profile and a 3 dB bandwidth of 17nm.}
\vspace*{1mm}
\label{Table:C}
\begin{threeparttable}
\begin{tabular}{|l|l|l|}
\hline
Wavelengths $\lambda_s$, $\lambda_i$ & Concurrence & Fidelity \\
\hline
1330 nm, 1290 nm & 0.9570 $\pm$ 0.0075 & 0.9779 $\pm$ 0.0037\\
\hline
1350 nm, 1270 nm & 0.9372 $\pm$ 0.0124 & 0.9658 $\pm$ 0.0063\\
\hline
1370 nm\tnote{*} & 0.9110 $\pm$ 0.0131 & 0.9546 $\pm$ 0.0065\\
\hline
\end{tabular}
\vspace{1mm}
\begin{tablenotes}
\footnotesize
\item[*] See caption of Fig.\ref{fig:QST}(d).
\end{tablenotes}
\end{threeparttable}
\end{table}

In summary, we have demonstrated a broadband polarization-entangled photon-pair source in PPSF by exploiting type-II SPDC at telecom O-band. The cw-pumped PPSF source emits highly time-correlated and high quality polarization-entangled biphotons of more than 130 nm bandwidth. The concurrence of polarization entanglement is measured to be greater than 0.91 even when the signal-idler is greater than 120 nm apart. The O-band PPSF source can be further optimized. For instance, the PPSF in the context of this letter has a non-zero chromatic dispersion ($|k_2|\sim2ps^2/km$) at the current QPM wavelength. The biphoton emission spectrum can be potentially broadened to be more than a few hundreds of nanometers if the QPM wavelength is tuned to be closer to the zero-dispersion wavelength of the PPSF. Moreover, using better components such as pump suppression filters and beam splitters with lower loss and better performance will surely result in higher entanglement quality as well.

This simple fiber-based source is compatible with existing fiber infrastructure. Operating near the zero-dispersion wavelength of silica fiber, the time-correlated and polarization entangled biphotons generated in this source are expected to be robust against chromatic dispersion and polarization mode dispersion in long-distance quantum fiber communication\cite{Jones2018, Brodsky2010, Shtaif2011}. Through further development, we anticipate that the source can be used in quantum applications such as wavelength-multiplexed quantum communication\cite{Wengerowsky2018, Zhu2019,Joshi2020} and high precision quantum sensing\cite{kaiser2018quantum, valencia2004distant, Riazi2019, nasr2003demonstration, Kalashnikov2016,riazi2019alignment}.

\medskip
\noindent\textbf{Funding.}
Natural Sciences and Engineering Research Council of Canada (RGPIN-2019-07019, RGPAS-2019-00113, CREATE-484907-16), Mitacs Globalink Research Award-Abroad, and US Army Award W911NF20-2-0242. The U.S. Government has a copyright license in this work pursuant to a Cooperative Research and Development Agreement with University of Toronto

\medskip
\noindent\textbf{Disclosures.} The authors declare no conflicts of interest.

% Bibliography
\bibliography{sample}

\begin{thebibliography}{10}
\newcommand{\enquote}[1]{``#1''}

\bibitem{slussarenko2019photonic}
S.~Slussarenko and G.~J. Pryde, {\protect\JournalTitle{Applied Physics
  Reviews}} \textbf{6}, 041303 (2019).

\bibitem{kwiat1995new}
P.~G. Kwiat, K.~Mattle, H.~Weinfurter, A.~Zeilinger, A.~V. Sergienko, and
  Y.~Shih, {\protect\JournalTitle{Physical Review Letters}} \textbf{75}, 4337
  (1995).

\bibitem{zhang2016engineering}
Y.~Zhang, F.~S. Roux, T.~Konrad, M.~Agnew, J.~Leach, and A.~Forbes,
  {\protect\JournalTitle{Science advances}} \textbf{2}, e1501165 (2016).

\bibitem{kaiser2018quantum}
F.~Kaiser, P.~Vergyris, D.~Aktas, C.~Babin, L.~Labont{\'e}, and S.~Tanzilli,
  {\protect\JournalTitle{Light: Science \& Applications}} \textbf{7}, 17163
  (2018).

\bibitem{nasr2003demonstration}
M.~B. Nasr, B.~E.~A. Saleh, A.~V. Sergienko, and M.~C. Teich,
  {\protect\JournalTitle{Phys. Rev. Lett.}} \textbf{91}, 083601 (2003).

\bibitem{lim2008broadband}
H.~C. Lim, A.~Yoshizawa, H.~Tsuchida, and K.~Kikuchi,
  {\protect\JournalTitle{Optics express}} \textbf{16}, 16052 (2008).

\bibitem{valencia2004distant}
A.~Valencia, G.~Scarcelli, and Y.~Shih, {\protect\JournalTitle{Applied Physics
  Letters}} \textbf{85}, 2655 (2004).

\bibitem{grieve2019characterizing}
J.~A. Grieve, Y.~Shi, H.~S. Poh, C.~Kurtsiefer, and A.~Ling,
  {\protect\JournalTitle{Applied Physics Letters}} \textbf{114}, 131106 (2019).

\bibitem{Zhong:09OE}
T.~Zhong, F.~N.~C. Wong, T.~D. Roberts, and P.~Battle,
  {\protect\JournalTitle{Opt. Express}} \textbf{17}, 12019 (2009).

\bibitem{Zhong2010}
T.~Zhong, X.~Hu, F.~N.~C. Wong, K.~K. Berggren, T.~D. Roberts, and P.~Battle,
  {\protect\JournalTitle{Optics Letters}} \textbf{35}, 1392 (2010).

\bibitem{Martin:09OE}
A.~Martin, V.~Cristofori, P.~Aboussouan, H.~Herrmann, W.~Sohler, D.~Ostrowsky,
  O.~Alibart, and S.~Tanzilli, {\protect\JournalTitle{Opt. Express}}
  \textbf{17}, 1033 (2009).

\bibitem{Martin2010}
A.~Martin, A.~Issautier, H.~Herrmann, W.~Sohler, D.~B. Ostrowsky, O.~Alibart,
  and S.~Tanzilli, {\protect\JournalTitle{New Journal of Physics}} \textbf{12},
  103005 (2010).

\bibitem{Liu2014}
M.~T. Liu and H.~C. Lim, {\protect\JournalTitle{Optics Express}} \textbf{22},
  23261 (2014).

\bibitem{Shi2020}
Y.~Shi, S.~M. Thar, H.~S. Poh, J.~A. Grieve, C.~Kurtsiefer, and A.~Ling,
  {\protect\JournalTitle{Applied Physics Letters}} \textbf{117}, 124002 (2020).

\bibitem{Xiang2020}
Z.-H. Xiang, J.~Huwer, J.~Skiba-Szymanska, R.~M. Stevenson, D.~J.~P. Ellis,
  I.~Farrer, M.~B. Ward, D.~A. Ritchie, and A.~J. Shields,
  {\protect\JournalTitle{Communications Physics}} \textbf{3} (2020).

\bibitem{Hall2009}
M.~A. Hall, J.~B. Altepeter, and P.~Kumar, {\protect\JournalTitle{Optics
  Express}} \textbf{17}, 14558 (2009).

\bibitem{nasr2008ultrabroadband}
M.~B. Nasr, S.~Carrasco, B.~E.~A. Saleh, A.~V. Sergienko, M.~C. Teich, J.~P.
  Torres, L.~Torner, D.~S. Hum, and M.~M. Fejer, {\protect\JournalTitle{Phys.
  Rev. Lett.}} \textbf{100}, 183601 (2008).

\bibitem{booth2011polarization}
M.~C. Booth, B.~E. Saleh, and M.~C. Teich, {\protect\JournalTitle{Optics
  Communications}} \textbf{284}, 2542 (2011).

\bibitem{Joshi2020}
S.~K. Joshi, D.~Aktas, S.~Wengerowsky, M.~Lon{\v{c}}ari{\'{c}}, S.~P. Neumann,
  B.~Liu, T.~Scheidl, G.~C. Lorenzo, {\v{Z}}.~Samec, L.~Kling, A.~Qiu,
  M.~Razavi, M.~Stip{\v{c}}evi{\'{c}}, J.~G. Rarity, and R.~Ursin,
  {\protect\JournalTitle{Science Advances}} \textbf{6}, eaba0959 (2020).

\bibitem{kim2006phase}
T.~Kim, M.~Fiorentino, and F.~N.~C. Wong, {\protect\JournalTitle{Phys. Rev. A}}
  \textbf{73}, 012316 (2006).

\bibitem{wootters1998entanglement}
W.~K. Wootters, {\protect\JournalTitle{Physical Review Letters}} \textbf{80},
  2245 (1998).

\bibitem{christ2013single}
A.~Christ, A.~Fedrizzi, H.~H{\"u}bel, T.~Jennewein, and C.~Silberhorn,
  \emph{Single-Photon Generation and Detection: Chapter 11. Parametric
  Down-Conversion}, vol.~45 (Elsevier Inc. Chapters, 2013).

\bibitem{zhu2012direct}
E.~Y. Zhu, Z.~Tang, L.~Qian, L.~G. Helt, M.~Liscidini, J.~E. Sipe, C.~Corbari,
  A.~Canagasabey, M.~Ibsen, and P.~G. Kazansky, {\protect\JournalTitle{Phys.
  Rev. Lett.}} \textbf{108}, 213902 (2012).

\bibitem{chen2017compensation}
C.~Chen, E.~Y. Zhu, A.~Riazi, A.~V. Gladyshev, C.~Corbari, M.~Ibsen, P.~G.
  Kazansky, and L.~Qian, {\protect\JournalTitle{Optics Express}} \textbf{25},
  22667 (2017).

\bibitem{zhu2013poled}
E.~Y. Zhu, Z.~Tang, L.~Qian, L.~G. Helt, M.~Liscidini, J.~E. Sipe, C.~Corbari,
  A.~Canagasabey, M.~Ibsen, and P.~G. Kazansky, {\protect\JournalTitle{Opt.
  Lett.}} \textbf{38}, 4397 (2013).

\bibitem{CorningSMF}
Corning, \emph{Corning SMF-28 Optical Fibers}.

\bibitem{Canagasabey2009}
A.~Canagasabey, C.~Corbari, A.~V. Gladyshev, F.~Liegeois, S.~Guillemet,
  Y.~Hernandez, M.~V. Yashkov, A.~Kosolapov, E.~M. Dianov, M.~Ibsen, and P.~G.
  Kazansky, {\protect\JournalTitle{Optics Letters}} \textbf{34}, 2483 (2009).

\bibitem{Zhu2010}
E.~Y. Zhu, L.~Qian, L.~G. Helt, M.~Liscidini, J.~E. Sipe, C.~Corbari,
  A.~Canagasabey, M.~Ibsen, and P.~G. Kazansky, {\protect\JournalTitle{Optics
  Letters}} \textbf{35}, 1530 (2010).

\bibitem{HOM1981}
C.~K. Hong, Z.~Y. Ou, and L.~Mandel, {\protect\JournalTitle{Phys. Rev. Lett.}}
  \textbf{59}, 2044 (1987).

\bibitem{branczyk2017hongoumandel}
A.~M. Brańczyk, \enquote{Hong-ou-mandel interference,}  (2017).

\bibitem{qstDFVJames2001PRA}
D.~F.~V. James, P.~G. Kwiat, W.~J. Munro, and A.~G. White,
  {\protect\JournalTitle{Phys. Rev. A}} \textbf{64}, 052312 (2001).

\bibitem{Jones2018}
D.~E. Jones, B.~T. Kirby, and M.~Brodsky, {\protect\JournalTitle{npj Quantum
  Information}} \textbf{4} (2018).

\bibitem{Brodsky2010}
M.~Brodsky, E.~C. George, C.~Antonelli, and M.~Shtaif,
  {\protect\JournalTitle{Optics Letters}} \textbf{36}, 43 (2010).

\bibitem{Shtaif2011}
M.~Shtaif, C.~Antonelli, and M.~Brodsky, {\protect\JournalTitle{Optics
  Express}} \textbf{19}, 1728 (2011).

\bibitem{Wengerowsky2018}
S.~Wengerowsky, S.~K. Joshi, F.~Steinlechner, H.~Hübel, and R.~Ursin,
  {\protect\JournalTitle{Nature}} \textbf{564}, 225 (2018).

\bibitem{Zhu2019}
E.~Y. Zhu, C.~Corbari, A.~Gladyshev, P.~G. Kazansky, H.-K. Lo, and L.~Qian,
  {\protect\JournalTitle{Journal of the Optical Society of America B}}
  \textbf{36}, B1 (2019).

\bibitem{Riazi2019}
A.~Riazi, C.~Chen, E.~Y. Zhu, A.~V. Gladyshev, P.~G. Kazansky, J.~E. Sipe, and
  L.~Qian, {\protect\JournalTitle{npj Quantum Information}} \textbf{5}, 1
  (2019).

\bibitem{Kalashnikov2016}
D.~A. Kalashnikov, A.~V. Paterova, S.~P. Kulik, and L.~A. Krivitsky,
  {\protect\JournalTitle{Nature Photonics}} \textbf{10}, 98 (2016).

\bibitem{riazi2019alignment}
A.~Riazi, E.~Y. Zhu, C.~Chen, A.~V. Gladyshev, P.~G. Kazansky, and L.~Qian,
  {\protect\JournalTitle{Optics letters}} \textbf{44}, 1484 (2019).

\end{thebibliography}

% Full bibliography added automatically for Optics Letters submissions; the following line will simply be ignored if submitting to other journals.
% Note that this extra page will not count against page length
\bibliographyfullrefs{sample}

%Manual citation list
%\begin{thebibliography}{1}
%\bibitem{Zhang:14}
%Y.~Zhang, S.~Qiao, L.~Sun, Q.~W. Shi, W.~Huang, %L.~Li, and Z.~Yang,
 % \enquote{Photoinduced active terahertz metamaterials with nanostructured
  %vanadium dioxide film deposited by sol-gel method,} Opt. Express \textbf{22},
  %11070--11078 (2014).
%\end{thebibliography}

\end{document}